\documentclass{article}
\usepackage{spconf,amsmath,graphicx}
\usepackage{amsfonts}
\usepackage{gensymb}
\usepackage{algorithmic}
\usepackage{graphicx}
\usepackage{textcomp}
\usepackage{xcolor}
\usepackage{hyperref}
\usepackage{subfigure}
\usepackage[bottom]{footmisc}

\title{soft-threshold attention based audio-visual speech enhancement network}
\name{Xinmeng Xu$^{1}$\quad Jianjun Hao$^{2*}$ \thanks{$^*$Corresponding author}}
\address{$^{1}$Electronic \& Elect. Engineering, Trinity College Dublin, Ireland\\
$^{2}$School of Foreign Languages, Hubei University of Chinese Medicine, P.R. China}
\begin{document}

%
\maketitle
\begin{abstract}
Audio-visual speech enhancement system is regarded to be one of the promising solutions for isolating and enhancing the speech of the desired speaker. Conventional methods focus on predicting clean speech spectrum via a naive convolution neural network based encoder-decoder architecture, and these methods a) are not adequate to use data fully and effectively, b) cannot process features selectively. To tackle these problems, this paper proposes a soft-threshold attention based convolution recurrent network for audio-visual speech enhancement, which a) applies a novel audio-visual fusion strategy that fuses audio and visual features layer by layer in encoding stage, and that feeds fused audio-visual features to each corresponding decoder layer, and more importantly, b) which introduces a soft-threshold attention applied on every decoder layers to select the informative modality softly. Experimental results illustrate that the proposed architecture obtains consistently better performance than recent models of both PESQ and STOI scores.

\end{abstract}
\section{Introduction}
\label{sec:intro}

Speech processing systems are commonly used in a variety of applications such as automatic speech recognition, speech synthesis, and speaker verification. Numerous speech processing devices (e.g. mobile communication systems and digital hearing aids systems) are often used in environments with high levels of ambient noise such as public places and cars in our daily life. Generally speaking, the presence of high-level noise interference, severely decrease perceptual quality and intelligibility of speech signal. Therefore, there is an urgent need for the development of speech enhancement algorithms which can automatically filter out noise signal and improve the effectiveness of speech processing systems.

Recently, many approaches are proposed to recover the clean speech of target speaker immersed in noisy environment, which can be roughly divided into two categories, i.e., audio-only speech enhancement (AO-SE)\cite{ref1, ref2} and audio-visual speech enhancement (AV-SE)\cite{ref4, ref5}. AO-SE approaches make assumptions on statistical properties of the involved signals\cite{ref14}, and aim to estimate target speech signals according to mathematically tractable criteria\cite{ref15}. Advanced AO-SE methods based on deep learning can predict target speech signal directly, but they tend to depart from the knowledge-based modelling. Compared with AO-SE approaches, AV-SE methods have achieved an improvement in the performance of intelligibility of speech enhancement due to the visual aspect which can recover some of the suppressed linguistic features when target speech is corrupted by noise interference\cite{ref21}. However, AV-SE model should be trained using data that are representative of settings in which they are deployed. In order to have robust performance in a wide variety of settings, very large AV datasets for training and testing need to be collected. Furthermore, AV-SE is inherently a multi-modal process, and it focuses not only on determining the parameters of a model, but also on the possible fusion architectures\cite{ref8}. Generally, a naive fusion strategy does not allow to control how the information from audio and the visual modalities is fused, as a consequence, one of the two modalities dominate the other.
\begin{figure*}[t]
  \centering
  \includegraphics[width=\linewidth]{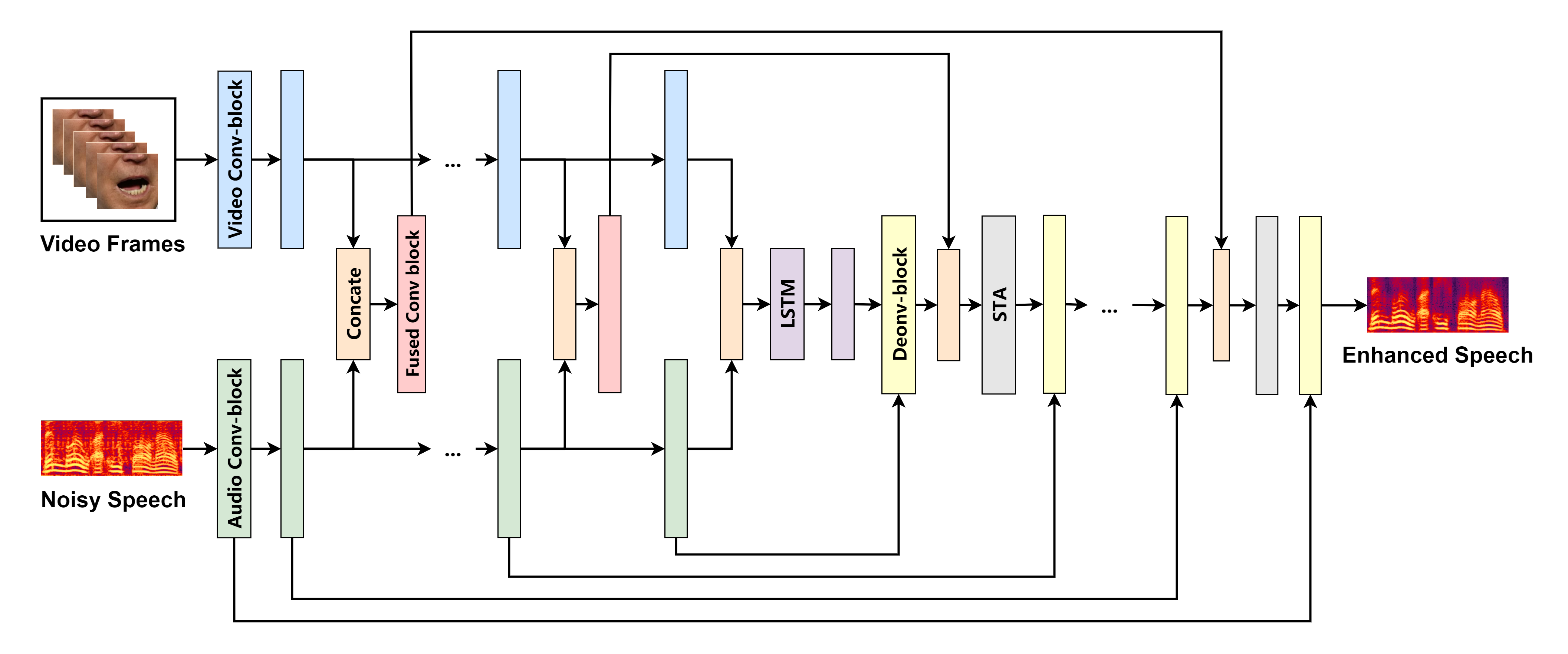}
  \caption{Schematic diagram of the proposed soft-threshold attention based CRN model. The STA denotes the soft-threshold attention unit.}
  \label{fig:0}
\end{figure*}

To overcome the aforementioned limitations, this paper proposes a Soft-threshold attention (STA) based Audio-visual Convolution Recurrent Neural Networks (AVCRN) for speech enhancement, which integrates the selected audio and visual cues into a unified network using multi-layer audio-visual fusion strategy. The proposed framework applies a STA inspired by soft thresholding algorithm \cite{ref10}, which has often been used as a key step in many signal denoising methods \cite{ref11}, and eliminates unimportant features \cite{zhao2019deep}. Moreover, the proposed model adopts the multi-layer audio and visual fusion strategy. in which the extracted audio and visual features are concatenated in every encoding layer. When two modalities in each layer are concatenated, the system applies them as an additional input via STA to feed the corresponding decoding layer. 

The rest of this paper is organized as follows: In Section 2, the proposed method is presented in detail. Section 3 is the dataset and experimental settings. Section 4 demonstrates the results and analysis, and a conclusion is shown in Section 5.

\section{Model Architecture}
\label{sec:format}

\subsection{Audio-visual CRN}
The diagram of proposed audio-visual CRN is demonstrated in Figure~\ref{fig:0}. This model following an encoder-decoder scheme, uses a series of downsampling and upsampling blocks to make its predictions, and consists of the encoder component, fusion component, and decoder component. 

The encoder component involves audio encoder and video encoder. As previous approaches in several CNNs based audio encoding models\cite{ref29, ref30, ref31}, the audio encoder is thus designed as a CNNs using the spectrogram as input. The video encoder part is used to process the input face embedding. In our approach, the video feature vectors and audio feature vectors take concatenation access at every step in the encoding stage, and the size of visual feature vectors after convolution layer has to be the same as the corresponding audio feature vectors, as is shown in Figure~\ref{fig:0}. 

Fusion component consists of audio-visual fusion process and audio-visual embedding process. Audio-visual fusion process usually designates a consolidated dimension to implement fusion, which combines the audio and visual streams in each layer directly and feeds the combination into several convolution layers. Audio-visual embedding which ﬂattens audio and visual streams from 3D to 1D, then concatenates both ﬂattened streams together, and ﬁnally feed the concatenated feature vector into two LSTM layers. Audio-visual embedding is a feature deeper fusion strategy, and the resulting vector is then to build decoder component. 

The decoder component, or named audio decoder, is made of deconvolutional layers. Because of the downsampling blocks, the model computes a number of higher level features on coarser time scales, and generates the audio-visual features by audio-visual fusion process in each level, which are concatenated with the local, high resolution features computed from the same level upsampling block. This concatenation results into multi-scale features for predictions.
\subsection{Soft-threshold attention}
\label{sec:pagestyle}
\begin{figure}
    \includegraphics[width=\linewidth]{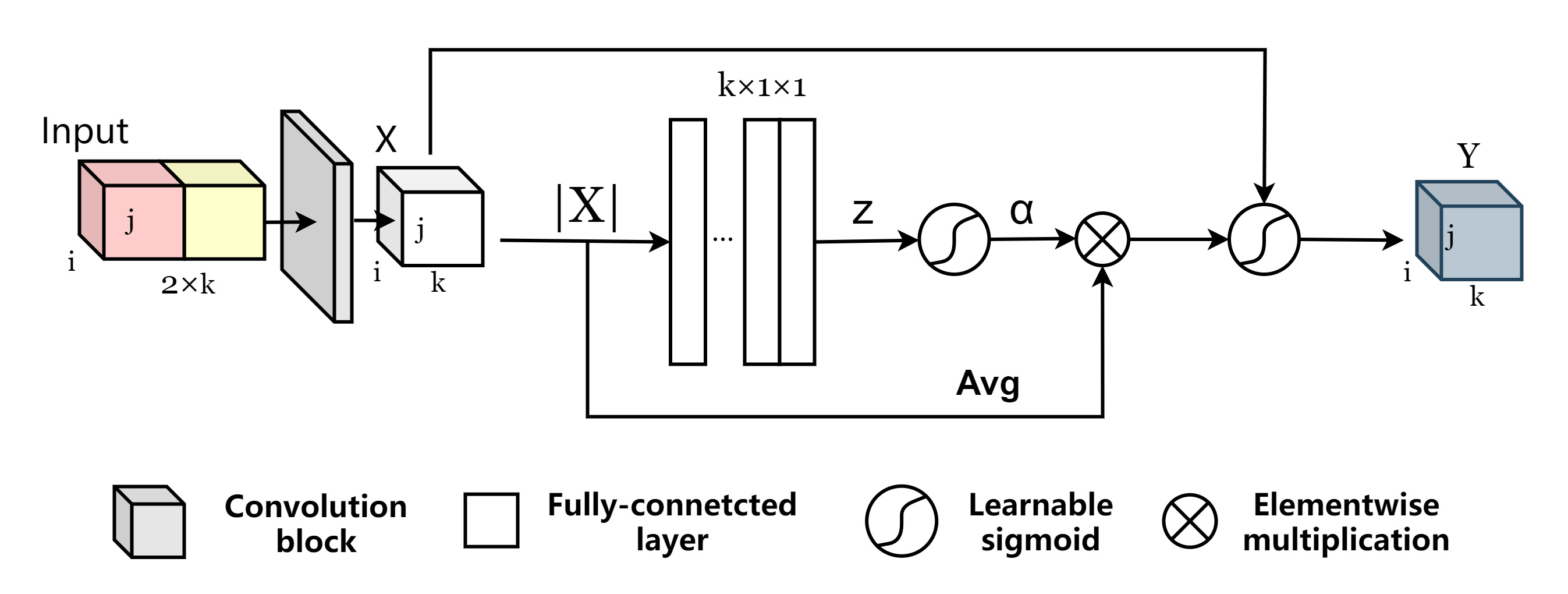}
    \centering
    \caption{The soft-threshold attention, where the $X_{i,j,k}$ is the feature map which generated by a convolution block with a concatenation input, $i$, $j$, and $k$ are the index of width, height and channel of the feature map $X$, $Y$ is output feature, which size is the same as $x$, and $z$, $\alpha$ are the indicators of the features maps to be used when determining threshold.}
    \label{fig2}
\end{figure}

In the proposed architecture, the potential unbalance caused by concatenation-based fusion easily happened on decoder blocks, when the concatenating features directly computed during contracting path with the same hierarchical level among the decoder blocks. Consequently, the proposed model use attention gates, as is shown in Figure~\ref{fig2}, to selectively filter out unimportant information using soft-thresholding algorithms.

Soft-thresholding is a kind of filter that can transform useful information to very positive or negative features and noise information to near-zero features. Deep learning enables the soft thresholding algorithm to be learned automatically by using a gradient decent algorithm , which is a promising way to eliminate noise-related information and construct highly discriminative features. The function of soft-thresholding can be expressed by
\begin{equation}
    Y=\begin{cases}
    X-\tau, & X > \tau \\
    0, & -\tau \leq X\leq \tau \\
    X+\tau, & X < -\tau \\
    \end{cases}
\label{eq1}
\end{equation}
where $X$ is the input feature, $Y$ is the output feature, and $\tau$ is the threshold. In addition, $X$ and $\tau$ are not independent variables where $\tau$ is non-negative, and their relation is expressed in Eq~\ref{eq3}.

The estimation of threshold is a set of deep learning blocks as is shown in Figure~\ref{fig2}. In the threshold estimating module, the feature map $X_{i,j,k}$, where $i$, $j$, and $k$ are the index of width, height and channel, is taken absolute value, and its dimension is reduced to 1D. The function of the following several fully-connected layers generates the attention mask\cite{ref32}, where the sigmoid function at the last layers scaled the attention mask from 0 to 1, which can be expressed by
\begin{equation}
    \alpha = \frac{1}{1 + e^{-z}}
\label{eq2}
\end{equation}
where $z$ is the output of fully-connected layers, and $\alpha$ is the attention mask. Finally, the threshold parameter $\tau$ can be used to determine the value of feature vectors, which are obtained by multiplying between the average value of $|X_{i, j, k}|$ and attention mask $\alpha$. The function of threshold parameter can be expressed by
\begin{equation}
    \tau = \alpha \times {\rm Avg}(|X_{i,j,k}|)
\label{eq3}
\end{equation}
where ${\rm Avg}(.)$ denotes the average pooling. Substitute Eq~\ref{eq2} and Eq~\ref{eq3} into Eq~\ref{eq1}, the output feature $Y_{i,j,k}$ can be obtained.

There are two advantages of STA: Firstly, it removes noise-related features from higher-level audio-visual fusion vectors. Secondly, it balances audio and visual modalities in the audio-visual fusion vector, and selectively take audio-visual features.

\section{Experimental setup}
\label{sec:print}
\subsection{Datasets}

The dataset used in the proposed model involves two publicly available audio-visual datasets: GRID\cite{ref33} and TCD-TIMIT\cite{ref34}, which are the two most commonly used databases in the area of audio-visual speech processing. GRID consists of video recordings where 18 male speakers and 16 female speakers pronounce 1000 sentences each. TCD-TIMIT consists of 32 male speakers and 30 female speakers with around 200 videos each.

The proposed model shuffles and splits the dataset to training, validation, and evaluation sets to 24300 (15 males, 12 females, 900 utterance each), 4400 (12 males, 10 females, 200 utterance each), and 1200 utterances (4 males, 4 females, 150 utterance each), respectively. The noise dataset contains 25.3 hours ambient noise categorized into 12 types:  room, car, instrument, engine, train, human chatting, air-brake, water, street, mic-noise, ring-bell, and music. 

Part of noise signals (23.9 hours) are conducted into both training set and validation set, but the rest are used to mix the evaluation set. The speech-noise mixtures in training and validation are generated by randomly selecting utterances from speech dataset and noise dataset and mixing them at random SNR between -10dB and 10dB. The evaluation set is generated SNR at 0dB and -5dB.

\subsection{Audio representation}
The audio representation is the transformed magnitude spectrograms in the log Mel-domain. The input audio signals are raw waveforms, and firstly are transformed to spectrograms using Short Time Fourier Transform (STFT) with Hanning window function, and 16 kHz resampling rate.  Each frame contains a window of 40 milliseconds, which equals 640 samples per frame and corresponds to the duration of a single video frame, and the frame shift is 160 samples (10 milliseconds). 

The transformed spectrograms are then converted to log Mel-scale spectrograms via Mel-scale filter banks. The resulting spectrogram have 80 Mel frequency bands from 0 to 8 kHz. The whole spectrograms are sliced into pieces of duration of 200 milliseconds corresponding to the length of 5 video frames, resulting in spectrograms of size 80$\times$20, representing 20 temporal samples, and 80 frequency bins in each sample.

\subsection{Video representation}
Visual representation is extracted from the input videos, and is re-sampled to 25 frames per second. Each video is divided into non-overlapping segments of 5 frames.

\section{Experiment Results}\label{sec4}
\renewcommand{\arraystretch}{1.2}
\begin{table*}[]
\caption{Models comparison in terms of STOI and PESQ scores, ``Speech'' interference denotes the background speech signal from unknown talker(s); ``Natural'' interference denotes the ambient non-speech noise.}
\centering
\begin{tabular}{|l|c|c|c|c|c|c|c|c|}
\hline
Evaluation metrics       & \multicolumn{4}{c|}{STOI (\%)}                                & \multicolumn{4}{c|}{PESQ}                                     \\ \hline
Test SNR                 & \multicolumn{2}{c|}{-5 dB}    & \multicolumn{2}{c|}{0 dB}     & \multicolumn{2}{c|}{-5 dB}    & \multicolumn{2}{c|}{0 dB}     \\ \hline
Interference                   & Speech        & Natural       & Speech        & Natural       & Speech        & Natural       & Speech        & Natural       \\ \hline
Unprocessed              & 57.8          & 51.4          & 64.7          & 62.6          & 1.59          & 1.03          & 1.66          & 1.24          \\ \hline
TCNN (Audio-only)        & 73.2          & 78.7          & 80.8          & 81.3          & 2.01          & 2.19          & 2.47          & 2.58          \\ \hline
Baseline & 77.9          & 81.3          & 88.6 & 87.9          & 2.41          & 2.35          & 2.77          & 2.94 \\ \hline
AV-CRN(proposed)                      & 80.7 & 82.7 & 88.4          & 89.3 & 2.61 & 2.72 & 2.84 & 2.92          \\ 
\quad+ Soft-threshold attention                      & \textbf{83.2} & \textbf{84.9} & \textbf{90.1}          & \textbf{92.5} & \textbf{2.81} & \textbf{2.94} & \textbf{3.04} & \textbf{3.11}          \\ \hline
\end{tabular}
\label{table:table2}
\end{table*}

\subsection{Competing models}
To evaluate the performance of the proposed approach, the comparisons are provided with several recently proposed speech enhancement algorithms. Specially, the evaluation methods are compared proposed model with TCNN model (an AO-SE approach), the AV-SE baseline system. Therefore, there are three networks have trained:
\begin{itemize}
    \item
    \textbf{TCNN}\cite{ref38}: Temporal convolutional neural network for real-time speech enhancement in the time domain.
    \item
    \textbf{Baseline}\cite{ref39}: A baseline work of visual speech enhancement.
    \item
    \textbf{STA-based CRN}: soft-threshold attention based convolution recurrent network for audio-visual speech enhancement.
\end{itemize}

\subsection{Results}
The results of the proposed network by using the following evaluation metrics: Short Term Objective Intelligibility (STOI) and Perceptual Evaluation of Speech Quality (PESQ). Each measurement compares the enhanced speech with clean reference of each of the test stimuli provided in the dataset. In addition, the proposed model has decomposed to two groups, AV-CRN model without STA \textit{i.e.} AV-CRN, and the complete form of proposed model, \textit{i.e.} AV-CRN + STA. \footnote{Speech samples are available at:
\scriptsize{\href{https://XinmengXu.github.io/AVSE/AVCRN.html}{\texttt{https://XinmengXu.github.\\io/AVSE/AVCRN.html}}}}

Table 1 demonstrates the improvement in the performance of network, as new component to the speech enhancement architecture, such as visual modality, multi-layer audio-visual feature fusion strategy, and finally the STA. There is an observation that the AV-SE baseline work outperforms TCNN, an end-to-end deep learning based AO-SE system, and the performance of AV-CRN model better than the baseline system. Hence the performance improvement from TCNN (AO-SE) to AV-CRN is primarily for two reasons: a) the addition of the visual modality, and b) the use of fusion technique named multi-layer audio-visual fusion strategy, instead of concatenating audio and visual modalities only once in the whole network. Finally, the results from Table 1 show that STA improves the performance of AV-CRN further. 

Table 2 demonstrates that our proposed approach produces state-of-the-art results in terms of speech quality metrics as is discussed above by comparing against the following three recently proposed methods that use deep neural networks to perform AV-SE on GRID dataset:
\begin{itemize}
    \item
    \textbf{Deep-learning-based AV-SE}\cite{ref40}: Deep-learning-based audio-visual speech enhancement in presence of Lombard effect
    \item
    \textbf{OVA approach}\cite{ref41}: A LSTM based AV-SE with mask estimation
    \item
    \textbf{L2L model}\cite{ref42}: A speaker independent audio-visual model for speech separation
\end{itemize}

The results where $\Delta$PESQ denotes PESQ improvement with AV-CRN + STA result in Table 1. Results for the competing methods are taken from the corresponding papers. Although the comparison results are for reference only, the proposed model demonstrates a robust performance in comparison with state-of-the-art results on the GRID AV-SE tasks.
\renewcommand{\arraystretch}{1.0}
\begin{table}[]
\caption{Performance comparison of proposed model with state-of-the-art result on GRID }
\centering
\begin{tabular}{|c|c|c|c|c|}
\hline
Test SNR                          & \multicolumn{2}{c|}{-5 dB} & \multicolumn{2}{c|}{0 dB} \\ \hline
Evaluation Metrics                   & \multicolumn{4}{c|}{$\Delta$PESQ}                              \\ \hline
Deep-learning-based AV-SE & \multicolumn{2}{c|}{1.09}  & \multicolumn{2}{c|}{0.77} \\ \hline
OVA Approach              & \multicolumn{2}{c|}{0.24}  & \multicolumn{2}{c|}{0.13} \\ \hline
L2L Model                 & \multicolumn{2}{c|}{0.28}  & \multicolumn{2}{c|}{0.16} \\ \hline
\end{tabular}
\label{tab2}
\end{table}

\section{Conclusion}\label{sec5}
This paper proposed an soft-threshold attention based convolution recurrent network for audio-visual speech enhancement. The multi-layer feature fusion strategy process a long temporal context by repeated downsampling and convolution of feature maps to combine both high-level and low-level features at different layer steps. In addition, STA is inspired by soft-thresholding algorithm, which can automatically select informative features, transfer them to very positive or negative features, and finally eliminate the rest of near-zero features. Results provided an illustration that the proposed model has better performance than some published state-of-the-art models on the GRID dataset.

\vfill\pagebreak
\bibliographystyle{IEEEbib}
\bibliography{refs}

\begin{thebibliography}{10}

\bibitem{ref1}
Li-Ping Yang and Qian-Jie Fu,
\newblock ``Spectral subtraction-based speech enhancement for cochlear implant
  patients in background noise,''
\newblock {\em The journal of the Acoustical Society of America}, vol. 117, no.
  3, pp. 1001--1004, 2005.

\bibitem{ref2}
K~Paliwal and Anjan Basu,
\newblock ``A speech enhancement method based on kalman filtering,''
\newblock in {\em ICASSP'87. IEEE International Conference on Acoustics,
  Speech, and Signal Processing}. IEEE, 1987, vol.~12, pp. 177--180.

\bibitem{ref4}
Jen-Cheng Hou, Syu-Siang Wang, Ying-Hui Lai, Yu~Tsao, Hsiu-Wen Chang, and
  Hsin-Min Wang,
\newblock ``Audio-visual speech enhancement using multimodal deep convolutional
  neural networks,''
\newblock {\em IEEE Transactions on Emerging Topics in Computational
  Intelligence}, vol. 2, no. 2, pp. 117--128, 2018.

\bibitem{ref5}
Ibrahim Almajai and Ben Milner,
\newblock ``Visually derived wiener filters for speech enhancement,''
\newblock {\em IEEE Transactions on Audio, Speech, and Language Processing},
  vol. 19, no. 6, pp. 1642--1651, 2010.

\bibitem{ref14}
Yariv Ephraim,
\newblock ``Statistical-model-based speech enhancement systems,''
\newblock {\em Proceedings of the IEEE}, vol. 80, no. 10, pp. 1526--1555, 1992.

\bibitem{ref15}
Afshin Rezayee and Saeed Gazor,
\newblock ``An adaptive klt approach for speech enhancement,''
\newblock {\em IEEE Transactions on Speech and Audio Processing}, vol. 9, no.
  2, pp. 87--95, 2001.

\bibitem{ref21}
Quentin Summerfield,
\newblock ``Use of visual information for phonetic perception,''
\newblock {\em Phonetica}, vol. 36, no. 4-5, pp. 314--331, 1979.

\bibitem{ref8}
Dhanesh Ramachandram and Graham~W Taylor,
\newblock ``Deep multimodal learning: A survey on recent advances and trends,''
\newblock {\em IEEE Signal Processing Magazine}, vol. 34, no. 6, pp. 96--108,
  2017.

\bibitem{ref10}
David~L Donoho,
\newblock ``De-noising by soft-thresholding,''
\newblock {\em IEEE transactions on information theory}, vol. 41, no. 3, pp.
  613--627, 1995.

\bibitem{ref11}
Minghang Zhao, Shisheng Zhong, Xuyun Fu, Baoping Tang, and Michael Pecht,
\newblock ``Deep residual shrinkage networks for fault diagnosis,''
\newblock {\em IEEE Transactions on Industrial Informatics}, vol. 16, no. 7,
  pp. 4681--4690, 2019.

\bibitem{zhao2019deep}
Minghang Zhao, Shisheng Zhong, Xuyun Fu, and Tang,
\newblock ``Deep residual shrinkage networks for fault diagnosis,''
\newblock {\em IEEE Transactions on Industrial Informatics}, vol. 16, no. 7,
  pp. 4681--4690, 2019.

\bibitem{ref29}
Szu-Wei Fu, Yu~Tsao, and Xugang Lu,
\newblock ``{SNR}-aware convolutional neural network modeling for speech
  enhancement.,''
\newblock in {\em Interspeech}, 2016, pp. 3768--3772.

\bibitem{ref30}
Tomas Kounovsky and Jiri Malek,
\newblock ``Single channel speech enhancement using convolutional neural
  network,''
\newblock in {\em 2017 IEEE International Workshop of Electronics, Control,
  Measurement, Signals and their Application to Mechatronics (ECMSM)}. IEEE,
  2017, pp. 1--5.

\bibitem{ref31}
Ke~Tan and DeLiang Wang,
\newblock ``A convolutional recurrent neural network for real-time speech
  enhancement.,''
\newblock in {\em Interspeech}, 2018, pp. 3229--3233.

\bibitem{ref32}
Jie Hu, Li~Shen, and Gang Sun,
\newblock ``Squeeze-and-excitation networks,''
\newblock in {\em Proceedings of the IEEE conference on computer vision and
  pattern recognition}, 2018, pp. 7132--7141.

\bibitem{ref33}
Martin Cooke, Jon Barker, Stuart Cunningham, and Xu~Shao,
\newblock ``An audio-visual corpus for speech perception and automatic speech
  recognition,''
\newblock {\em The Journal of the Acoustical Society of America}, vol. 120, no.
  5, pp. 2421--2424, 2006.

\bibitem{ref34}
Naomi Harte and Eoin Gillen,
\newblock ``{TCD-TIMIT}: An audio-visual corpus of continuous speech,''
\newblock {\em IEEE Transactions on Multimedia}, vol. 17, no. 5, pp. 603--615,
  2015.

\bibitem{ref38}
Ashutosh Pandey and DeLiang Wang,
\newblock ``{TCNN}: Temporal convolutional neural network for real-time speech
  enhancement in the time domain,''
\newblock in {\em ICASSP 2019-2019 IEEE International Conference on Acoustics,
  Speech and Signal Processing (ICASSP)}. IEEE, 2019, pp. 6875--6879.

\bibitem{ref39}
Aviv Gabbay, Asaph Shamir, and Shmuel Peleg,
\newblock ``Visual speech enhancement,''
\newblock {\em Interspeech}, pp. 1170--1174, 2018.

\bibitem{ref40}
Daniel Michelsanti, Zheng-Hua Tan, Sigurdur Sigurdsson, and Jesper Jensen,
\newblock ``Deep-learning-based audio-visual speech enhancement in presence of
  lombard effect,''
\newblock {\em Speech Communication}, vol. 115, pp. 38--50, 2019.

\bibitem{ref41}
Wupeng Wang, Chao Xing, Dong Wang, Xiao Chen, and Fengyu Sun,
\newblock ``A robust audio-visual speech enhancement model,''
\newblock in {\em ICASSP 2020-2020 IEEE International Conference on Acoustics,
  Speech and Signal Processing (ICASSP)}. IEEE, 2020, pp. 7529--7533.

\bibitem{ref42}
Ariel Ephrat, Inbar Mosseri, Oran Lang, Tali Dekel, Kevin Wilson, Avinatan
  Hassidim, William~T Freeman, and Michael Rubinstein,
\newblock ``Looking to listen at the cocktail party: A speaker-independent
  audio-visual model for speech separation,''
\newblock {\em ACM Transactions on Graphics}, 2018.

\end{thebibliography}

\end{document}